\documentclass[journal]{IEEEtran}

\usepackage{graphicx}
\usepackage{amssymb}
\usepackage{lineno}
\usepackage{enumerate}
\usepackage{enumitem}
\usepackage{amsthm}
\usepackage{multirow}
\usepackage[latin1]{inputenc}

\usepackage{epsfig}
\usepackage{algorithm}
\usepackage[noend]{algorithmic}
\algsetup{indent=2em}
\usepackage{epsfig}
\usepackage{url}
\usepackage{fancyhdr}
\usepackage{tabularx}
\usepackage{graphics}
\usepackage{float}
\usepackage{lscape}
\usepackage{rotating}
\usepackage{multicol}
\usepackage{pifont}
\newtheorem{definition}{Definition}

\usepackage[cmex10]{amsmath}
\usepackage{array}
\usepackage{url}

\begin{document}

\newpage

\title{ND-Tree-based update: a Fast Algorithm for the Dynamic Non-Dominance Problem 
}

\author{Andrzej~Jaszkiewicz,
        Thibaut~Lust
        
\thanks{A. Jaszkiewicz is with Poznan University of Technology, Faculty of Computing, Institute of Computing Science, ul. Piotrowo 2, 60-965 Poznan, Poland, e-mail: andrzej.jaszkiewicz@put.poznan.pl.}
\thanks{T. Lust is with Sorbonne Universit\'es, UPMC, Universit\'e Paris 06, CNRS, LIP6, UMR 7606, F-75005, Paris, France, e-mail: thibaut.lust@lip6.fr.}}

\maketitle

\begin{abstract}
In this paper we propose a new method called ND-Tree-based update (or shortly ND-Tree) for the dynamic non-dominance problem, i.e. the problem of online update of a Pareto archive composed of mutually non-dominated points. It uses a new ND-Tree data structure in which each node represents a subset of points contained in a hyperrectangle defined by its local approximate ideal and nadir points. By building subsets containing points located close in the objective space and using basic properties of the local ideal and nadir points we can efficiently avoid searching many branches in the tree. ND-Tree may be used in multiobjective evolutionary algorithms and other multiobjective metaheuristics to update an archive of potentially non-dominated points. We prove that the proposed algorithm has sub-linear time complexity under mild assumptions. We experimentally compare ND-Tree to the simple list, Quad-tree, and M-Front methods using artificial and realistic benchmarks with up to 10 objectives and show that with this new method substantial reduction of the number of point comparisons and computational time can be obtained. Furthermore, we apply the method to the non-dominated sorting problem showing that it is highly competitive to some recently proposed algorithms dedicated to this problem.
\end{abstract}


\IEEEpeerreviewmaketitle

\section{Introduction}

\IEEEPARstart{I}{n} this paper we consider the dynamic non-dominance problem~\cite{Schutze:2003:NDS:1760102.1760145}, i.e. the problem of online update of a Pareto archive with a new candidate point. The Pareto archive is composed of mutually non-dominated points and this property must remain fulfilled following the addition of the new point.  

The dynamic non-dominance problem is typically used in multiobjective evolutionary algorithms (MOEAs) and more generally in other multiobjective metaheuristics (MOMHs), whose goal is to generate a good approximation of the Pareto front.  Many MOEAs and other MOMHs use an external archive of potentially non-dominated points, i.e. a Pareto archive containing points not dominated by any other points generated so far, see e.g. \cite{zhang2007moea,Zitzler01,Coello2004,Agrawal2008,Li2016,Cai2015ExternalArchive,Yang2013, Tang2013}. We consider here MOEAs that generate iteratively new candidate points and use them immediately to update a Pareto archive. Updating a Pareto archive with a new point $y$ means that:
\begin{itemize}
\item $y$ is added to the Pareto archive if it is non-dominated w.r.t. any point in the Pareto archive,
\item all points dominated by $y$ are removed from the Pareto archive.
\end{itemize}

The time needed to update a Pareto archive, in general, increases with a growing number of objectives and a growing number of points. In some cases it may become a crucial part of the total running time of a MOEA. The simplest data structure for storing a Pareto archive is a plain list. When a new point $y$ is added, $y$ is compared to all points in the Pareto archive until either all points are checked or a point dominating $y$ is found. In order to speed up the process of updating a Pareto archive some authors proposed the use of specialized data structures and algorithms, e.g. Quad-tree~\cite{Mostaghim2004}. However, the results of computational experiments reported in literature are not conclusive and in some cases such data structures may in fact increase the update time compared to the simple list.

A frequently used approach allowing reduction of the time needed to update a Pareto archive is the use of bounded archives~\cite{Fieldsend2003} where the number of points is limited and some potentially non-dominated points are discarded. Please note, however, that such an approach always reduces the quality of the archive. In particular, one of the discarded points could be the one that would be selected by the decision maker if the full archive was known. Bounded archives may be especially disadvantageous in the case of many-objective problems, since with a growing number of dimensions it is more and more difficult to represent a large set with a smaller sample of points. The use of bounded archives may also lead to some technical difficulties in MOEAs (see \cite{Fieldsend2003}). Summarizing, if an unbounded archive can be efficiently managed and updated, it is advantageous to use this kind of archive.

In this paper, our contribution is fourfold: firstly, we propose a new method, called ND-Tree-based update, for the dynamic non-dominance problem. The method is based on a dynamic division of the objective space into hyperrectangles, which allows to avoid many comparisons of objective function values. Secondly, we show that the new method has sub-linear time complexity under mild assumptions. Thirdly, a thorough experimental study on different types of artificial and realistic sets shows that we can obtain substantial computational time reductions compared to state-of-the-art methods. Finally, we apply ND-Tree-based update to the non-dominated sorting problem obtaining promising results in comparison to some recently proposed dedicated algorithms.

The remainder of the paper is organized as follows. Basic definitions related to multiobjective optimization are given in Section II. In Section III, we present a state of the art of the methods used for online updating a Pareto archive. The main contribution of the paper, i.e. ND-Tree-based update method is described in Section IV. Computational experiments are reported and discussed in Section V. In sections VI, ND-Tree-based is applied to the non-dominated sorting problem.

\section{Basic definitions}

\subsection{Multiobjective optimization}

We consider a general multiobjective optimization (MO) problem with a feasible set of solutions $\mathcal{X}$ and $p$ objective functions $y_k(x)$ to minimize. The image of the feasible set in the objective space is a set of points $\mathcal{Y}=y(\mathcal{X})$ where $y(x)= (y_1(x),y_2(x),\ldots,y_p(x))$.  

In MO, points are usually compared according to the \emph{Pareto dominance relation}:

\begin{definition}
Pareto dominance relation: we say that a point $u=(u_1,...,u_p)$ \emph{dominates} a point $v=(v_1,...,v_p)$ if, and only if, $u_k \leq v_k\  \forall \, k \in \{1,\ldots,p\} \wedge \exists \, k \in \{1,\ldots,p\}: u_k < v_k$. We denote this relation by $u \succ v$. 
\end{definition}

\begin{definition}
Non-dominated point: a point $y^*$ corresponding to a feasible solution is called \emph{non-dominated} if there does not exist any other point $y \in \mathcal{Y}$ such that $y \succ y^*$. The set $\mathcal{Y}_N$ of all non-dominated points is called \emph{Pareto front}.
\end{definition}

\begin{definition}
Coverage relation: we say that a point $u$ \emph{covers} a point $v$ if $u \succ v$ or $u = v$. We denote this relation by $u \succeq v$. 
\end{definition}

Please note that coverage relation is sometimes referred to as weak dominance~\cite{Brockhoff2010}. 

\begin{definition}
Mutually non-dominated relation: we say that two points are \textit{mutually non-dominated} or \textit{non-dominated w.r.t. each other} if neither of the two points covers the other one. 
\end{definition}

\begin{definition}
Pareto archive ($Y_N$): set of points such that any pair of points in the set are mutually non-dominated, i.e $\forall y \in Y_N, \nexists y' \in Y_N \, | \, y' \succeq y$. 
\end{definition}

In the context of MOEAs, the Pareto archive contains the mutually non-dominated points generated so far (i.e. at a given iteration of a MOEA) that approximates the Pareto front $\mathcal{Y}_N$. In other words $Y_N$ contains points that are potentially non-dominated at a given iteration of the MOEA.

Please note that in MOEAs not only points but also representations of solutions are preserved in the Pareto archive, but the above definition is sufficient for the purpose of this paper. 

The new method ND-Tree-based update is based on the (approximate) local ideal and nadir points that we define below. 

\begin{definition}
The local ideal point of a subset $\mathcal{S} \subseteq Y_N$ denoted as $z^*(\mathcal{S})$ is the point in the objective space composed of the best coordinates of all points belonging to $\mathcal{S}$, i.e. $z^*_k(\mathcal{S}) =  \underset {y \in \mathcal{S}} \min \{ y_k \}, \forall \, k \in \{1,\ldots,p\}$. A point $\widehat{z}^*(\mathcal{S})$ such that $\widehat{z}^*(\mathcal{S}) \succeq z^*(\mathcal{S})$ will be called \emph{Approximate local ideal point}.
\end{definition}

Naturally, the (approximate) local ideal point covers all points in $\mathcal{S}$.

\begin{definition}
The local nadir point of a subset $\mathcal{S} \subseteq Y_N$ denoted as $z_*(\mathcal{S})$ is the point in the objective space composed of the worst coordinates of all points belonging to $\mathcal{S}$, i.e. $z_{*k}(\mathcal{S}) =  \underset {y \in \mathcal{S}} \max \{ y_k \}, \forall \, k \in \{1,\ldots,p\}$. A point $\widehat{z}_*(\mathcal{S})$ such that $z_*(\mathcal{S}) \succeq \widehat{z}_*(\mathcal{S})$ will be called \emph{Approximate local nadir point}.
\end{definition}

Naturally, the (approximate) local nadir point is covered by all points in $\mathcal{S}$.

\subsection{Dynamic non-dominance problem}

The problem of updating a Pareto archive (also called \emph{non-dominance problem}), can be divided into two classes: the \emph{static} non-dominance problem is to find the set of non-dominated points $Y_N$ among a set of points $Y$. The other class is the \emph{dynamic} non-dominance problem~\cite{Schutze:2003:NDS:1760102.1760145} that typically occurs in MOEAs. 
We formally define this problem as follows. Consider a candidate point $y$ and a Pareto archive $Y_N$. 
The problem is to update $Y_N$ with $y$ and consists in the following operations. If $y$ is covered by at least one point in $Y_N$, $y$ is discarded and $Y_N$ remains unchanged. Otherwise, $y$ is added to $Y_N$. Moreover, if some points in $Y_N$ are dominated by $y$, all these points are removed from $Y_N$, in order to keep only mutually non-dominated points (see Algorithm~\ref{algoDynamicNonDominance}). 

\floatname{algorithm}{Algorithm}
\begin{algorithm}[!ht]
\caption{\texttt{DynamicNonDominance}}\label{algoDynamicNonDominance}
\begin{algorithmic}
\STATE Parameter $\updownarrow$: A Pareto archive $Y_N$ 
\STATE Parameter $\downarrow$: New candidate point $y$

\vspace*{1\baselineskip}

\IF{($\nexists \,y' \in Y_N \ |\  y' \succeq y$)}
\STATE $Y_N \leftarrow Y_N \cup \{y\}$
\FORALL{($y' \in Y_N \,|\, y \succ y'$)}
\STATE $Y_N \leftarrow Y_N  \backslash \{y'\}$
\ENDFOR
\ENDIF

\end{algorithmic}
\end{algorithm}

In this work we consider only the dynamic non-dominance problem. Note that in general static problems may be solved more effectively than their dynamic counterparts since they have access to richer information. Indeed, some efficient algorithms for static non-dominance problem have been proposed, see~\cite{Kung1975,Gabow1984,Preparata1985,Gupta1997}. 

MOEAs and other MOMHs usually update the Pareto archive using the dynamic version of the non-dominance problem, i.e. the Pareto archive is updated with each newly generated candidate point. In some cases it could be possible to store all candidate points and then solve the static non-dominance problem. The latter approach has, however, some disadvantages:
\begin{itemize}
\item MOEAs need to store not only points in the objective space but also full representations of solutions in the decision space. Thus, storing all candidate points with corresponding solutions may be very memory consuming.
\item Some MOEAs use the Pareto archive during the run of the algorithm, i.e. Pareto archive is not just the final output of the algorithm. For example in \cite{Deb:2005:EED:1109044.1109049}, one of the parents is selected from the Pareto archive. In \cite{Cai2015ExternalArchive} the success of adding a new point to the Pareto archive influences the probability of selecting weight vectors in further iterations. The same applies to other MOMHs as well. For example, the Pareto local search (PLS) method~\cite{Paquete04} works directly with the Pareto archive and searches neighborhood of each solution from the archive. In such methods, computation of the Pareto archive cannot be postponed till the end of the algorithm.
\end{itemize}

Note that as suggested in~\cite{Drozdik2015} the dynamic non-dominance problem may also be used to speed up the non-dominated sorting procedure used in many MOEAs. As the Pareto archive contains all non-dominated points generated so far the first front is immediately known and the non-dominated sorting may be applied only to the subset of dominated points. Using this technique, Drozd\'ik \emph{et al.} showed that their new method called M-Front could obtain better performance than Deb's fast nondominated sorting~\cite{Deb00} and Jensen-Fortin's algorithm~\cite{Jensen03,Fortin13}, one of the fastest non-dominated sorting algorithms.

\section{State of the art}

We present here a number of methods for the dynamic non-dominance problem proposed in literature and used in the comparative experiment. This review is not supposed to be exhaustive. Other methods can be found in~\cite{Schutze03,Fieldsend03} and reviews in~\cite{Altwaijry2012,Mostaghim2002}. We describe linear list, Quad-tree and one recent method, M-Front~\cite{Drozdik2015}.

\subsection{Linear List}

\subsubsection{General case}

In this structure, a new point is compared to all points in the list until a covering point is found or all points are checked. The point is only added if it is non-dominated w.r.t. all points in the list, that is in the worst case we need to browse the whole list before adding a point. The complexity in terms of number of points comparison is thus in $\mathcal{O}(N)$ with $N$ the size in the list. 

\subsubsection{Biobjective case: sorted list}

When only two objectives are considered, we can use the following specific property: if we sort the list according to one objective (let's say the first), the non-dominated list is also sorted according to the second objective. Therefore, roughly speaking, updating the list can be efficiently done in the following way.
We first determine the potential position $i$ of the new candidate point in the sorted list according to its value of the first objective, with a binary search. If the new point is not dominated by the preceding one in the list (if there is one), the new point is not dominated and can be inserted at position $i$. If the new point has been added, we need to check if there are some dominated points: we browse the next points in the list, until a point is found that has a better evaluation according to the second objective. All the points found that have a worse evaluation according to the second objective have to be removed since they are dominated by the new point. 

The worst-case complexity is still in $\mathcal{O}(N)$ since it can happen that a new point has to be compared to all the other points (in the special case where we add a new point in the first position and all the points in the sorted list are dominated by this new point). But on average, experiments show that the behavior of this structure for handling biobjective updating problems is much better than the simple list. 

The algorithm of this method is given in Algorithm~\ref{algoSortedList} (for the sake of clarity, we only present the case where the candidate point $y$ has a distinct value for the first objective compared to all the other points in the archive $Y_N$).

\floatname{algorithm}{Algorithm}
\begin{algorithm}[!ht]
\caption{\texttt{Sorted list}}\label{algoSortedList}
\begin{algorithmic}
\STATE Parameter $\updownarrow$: A biobjective Pareto archive $Y_N$ 
\STATE Parameter $\downarrow$: New candidate point $y$
\vspace*{1\baselineskip}
\IF{$Y_N=\emptyset$}
\STATE $Y_N \leftarrow Y_N \cup \{y\}$
\ELSE{}
\STATE -$\,$-$|$ Looking for the position $i$ of $y$ in $Y_N$
\STATE $i \leftarrow$ \texttt{BinarySearch}$(Y_N,y_1)$
\IF{($i=0$) \OR $(y_2<y^{(i-1)}_2)$}
\STATE -$\,$-$|$ $y$ is added at position $i$ in $Y_N$
\STATE \texttt{Insert}$(Y_N,y,i)$
\STATE $j\leftarrow i+1$
\WHILE{$(j < |Y_N|)$ \AND $(y_2 \leq y^j_2)$}
\STATE -$\,$-$|$ $y^j$ is dominated
\STATE $Y_N \leftarrow Y_N \backslash \{y^j\}$
\STATE $j \leftarrow j+1$
\ENDWHILE
\ENDIF
\ENDIF
\end{algorithmic}
\end{algorithm}

\subsection{Quad-tree}
The use of Quad-tree for storing potentially non-dominated points was proposed by Habenicht ~\cite{Habenicht1983} and further developed by Sun and Steuer ~\cite{Sun1996} and Mostaghim and Teich ~\cite{Mostaghim2004}. In Quad-tree, points are located in both internal nodes and leaves. Each node may have $p^2$ children corresponding to each possible combination of results of comparisons on each objective where a point can either be better or not worse. In the case of  mutually non-dominated points $p^2-2$ children are possible since the combinations corresponding to dominating or covered points are not used.
Quad-tree allows for a fast checking if a new point is dominated or covered. A weak point of this data structure is that when an existing point is removed its whole sub-tree has to be re-inserted to the structure. Thus, removal of a dominated point is in general costly. 

\subsection{M-Front}

M-Front has been proposed relatively recently by Drozd\'ik \emph{et al.} ~\cite{Drozdik2015}. The idea of of M-Front is as follows. Assume that in addition to the new point $y$ a reference point $ref$ relatively close to $y$ and belonging to the Pareto archive $Y_N$ is known. The authors define two sets:

{\small
\[RS_U(y, ref)= \{z \in Y_N \mid  \exists k \in \{1,\ldots,p\}: z_k \geq ref_k \, \land \, z_k \leq y_k \}\]
\[RS_L(y, ref)= \{z \in Y_N \mid  \exists k \in \{1,\ldots,p\}: z_k \geq y_k \, \land  \,  z_k \leq ref_k \}\]}
and prove that if a point $z \in Y_N$ is dominated by $y$ then it belongs to $RS_L(y, ref)$ and if $z$ dominates $y$ then it belongs to $RS_U(y, ref)$. Thus, it is sufficient to compare the new points to sets $RS_L(y, ref)$ and $RS_U(y, ref)$ only. To find all points with objective values in a certain interval M-Front uses additional indexes one for each objective. Each index sorts the Pareto archive according to one objective.

To find a reference point close to $y$, M-Front uses the k-d tree data structure. The k-d tree is a binary tree, in which each intermediate node divides the space into two parts based on a value of one objective. While going down the tree the algorithm cycles over particular objectives, selecting one objective for each level. Drozd\'ik \emph{et al.}~\cite{Drozdik2015} suggest to store references to points in leaf nodes only, while intermediate nodes keep only split values. 

\section{ND-Tree-based update}

\subsection{Presentation}

In this section we present the main contribution of the paper. The new method for updating a Pareto archive is based on the idea of recursive division of archive $Y_N$ into subsets contained in different hyperrectangles. This division allows to considerably reduce the number of comparisons to be made. 

More precisely, consider a subset $\mathcal{S} \subseteq Y_N$ composed of mutually non-dominated points and a new candidate point $y$. Assume that some approximate local ideal $\widehat{z}^*(\mathcal{S})$ and approximate local nadir points $\widehat{z}_*(\mathcal{S})$ of $\mathcal{S}$ are known. In other words, all points in $\mathcal{S}$ are contained in the axes-parallel hyperrectangle defined by $\widehat{z}^*(\mathcal{S})$ and  $\widehat{z}_*(\mathcal{S})$. 

We can define the following simple  properties that allow to compare a new point $y$ to the whole set $\mathcal{S}$:
\begin{enumerate}[label=\textbf{Property} \textbf{\arabic*}.,itemindent=*]
\item If $y$ is covered by $\widehat{z}_*(\mathcal{S})$, then $y$ is covered by each point in $\mathcal{S}$ and thus can be rejected. This property is a straightforward consequence of the transitivity of the coverage relation.
\item If $y$ covers $\widehat{z}^*(\mathcal{S})$, then each point in $\mathcal{S}$ is covered by $y$. This property is also a straightforward consequence of the transitivity of the coverage relation.
\item	If $y$ is non-dominated w.r.t. both $\widehat{z}_*(\mathcal{S})$ and $\widehat{z}^*(\mathcal{S})$, then $y$ is non-dominated w.r.t. each point in  $\mathcal{S}$. \begin{proof} If $y$ is non-dominated w.r.t. $\widehat{z}_*(\mathcal{S})$ then there is at least one objective on which $y$ is worse than $\widehat{z}_*(\mathcal{S})$ and thus worse than each point in $\mathcal{S}$. If $y$ is non-dominated w.r.t. $\widehat{z}^*(\mathcal{S})$ then there is at least one objective on which $y$ is better than $\widehat{z}^*(\mathcal{S})$ and thus better than each point in $\mathcal{S}$. So, there is at least one objective on which $y$ is better and at least one objective on which $y$ is worse than each point in $\mathcal{S}$.\end{proof}
\end{enumerate}

If none of the above properties holds, i.e. $y$ is neither covered by $\widehat{z}_*(\mathcal{S})$, does not cover $\widehat{z}^*(\mathcal{S})$, nor is non-dominated w.r.t. both $\widehat{z}_*(\mathcal{S})$ and $\widehat{z}^*(\mathcal{S})$, then all situations are possible, i.e. $y$ may either be non-dominated w.r.t. all points in $\mathcal{S}$, covered by some points in $\mathcal{S}$ or dominate some points in $\mathcal{S}$. This can be illustrated by showing examples of each of the situations. Consider for example a set $\mathcal{S} = \{(1, 1, 1), (0, 2, 2), (2, 2, 0)\}$ with $\widehat{z}^*(\mathcal{S})=z^*(\mathcal{S}) = (0, 1, 0)$ and $\widehat{z}_*(\mathcal{S})=z_*(\mathcal{S}) = (2, 2, 2)$. A new point $(1, 1, 0)$ dominates a point in $\mathcal{S}$, a new point (1, 1, 2) is dominated (thus covered) by a point in $\mathcal{S}$, and points $(0, 3, 0)$ and $(2, 0, 1)$ are non-dominated w.r.t. all points in $\mathcal{S}$.

\begin{figure}
  \centering
  \includegraphics[scale=0.65,angle=0]{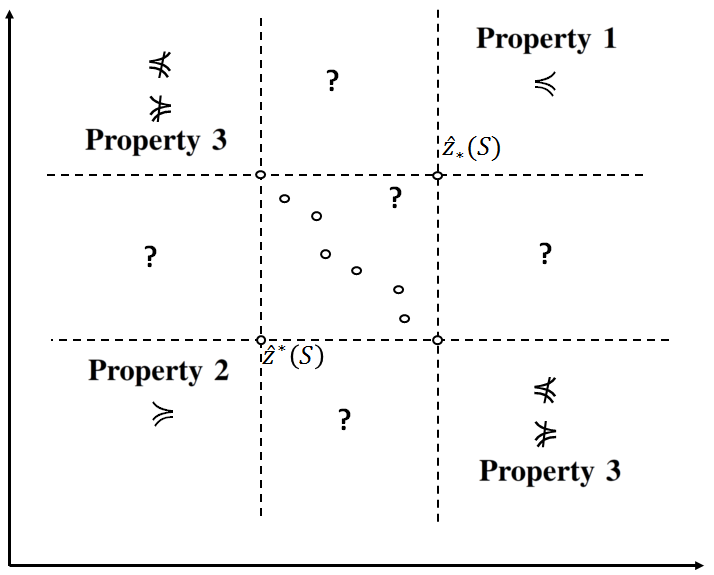}
   \caption{Comparison of a new point to all points in set $\mathcal{S}$ based on comparisons to $\widehat{z}^*(\mathcal{S})$ and $\widehat{z}_*(\mathcal{S})$ only.}
\label{fig:properties}
\end{figure}

The properties are graphically illustrated for the biobjective case in Figure~\ref{fig:properties}. As can be seen in this figure, in the biobjective case, if $y$ is covered by $\widehat{z}^*(\mathcal{S})$ and $y$ is non-dominated w.r.t. $\widehat{z}_*(\mathcal{S})$ then $y$ is dominated by at least one point in $\mathcal{S}$. Note, however, that this does not hold in the case of three and more objectives as shown in the above example - the point $(0,3,0)$ is covered by $\widehat{z}^*(\mathcal{S})=(0,1,0)$, non-dominated w.r.t. $\widehat{z}_*(\mathcal{S})=(2,2,2)$ and $(0,3,0)$ is not dominated by any points in $\mathcal{S}$.

In fact it is possible to distinguish more specific situations when none of the three properties hold, e.g. a situation when a new point may be covered but cannot dominate any point, but since we do not distinguish them in the proposed algorithm we do not define them formally.

The above properties allow in some cases to quickly compare a new candidate point $y$ to all points in set $\mathcal{S}$ without the need for further comparisons to individual points belonging to $\mathcal{S}$. Such further comparisons are necessary only if none of the three properties hold. Intuitively, the closer the approximate local ideal and nadir points the more likely it is that further comparisons can be avoided. To obtain close approximate local ideal and nadir points we should:
\begin{itemize}
\item Split the whole set of non-dominated points into subsets of points located close in the objective space.
\item Have good approximations of the exact local ideal and nadir points. On the other hand calculation of the exact points may be computationally demanding and a reasonable approximation may assure the best overall efficiency.
\end{itemize}

Based on these properties, we can now define the ND-Tree data structure. 

\begin{definition}
ND-Tree data structure is a tree with the following properties:
\begin{enumerate}
\item With each node $n$ is associated a set of points $\mathcal{S}(n)$.
\item Each leaf node contains a list $\mathcal{L}(n)$ of points and $\mathcal{S}(n) = \mathcal{L}(n)$.
\item For each internal node $n$, $\mathcal{S}(n)$ is the union of disjoint sets associated with all children of $n$.
\item Each node $n$ stores an approximate ideal point $\widehat{z^*}(\mathcal{S}(n))$ and approximate nadir point $\widehat{z_*}(\mathcal{S}(n))$.
\item If $n'$ is a child of $n$, then $\widehat{z^*}(\mathcal{S}(n)) \succeq \widehat{z^*}(\mathcal{S}(n'))$ and $\widehat{z_*}(\mathcal{S}(n')) \succeq \widehat{z_*}(\mathcal{S}(n))$.
\end{enumerate}

\end{definition}

The algorithm for updating a Pareto archive with ND-Tree is given in Algorithm~\ref{algoUpdate}. The idea of the algorithm is as follows. We start by checking if the new point $y$ is covered or non-dominated w.r.t. all points in $Y_N$ by going through the nodes of ND-Tree and skipping children (and thus their sub-trees) for which \textbf{Property 3} holds. This procedure is presented in Algorithm~\ref{algoUpdateNode}.

The new point is first compared to the approximate ideal point $(\widehat{z^*}(\mathcal{S}(n))$ and nadir point ($\widehat{z_*}(\mathcal{S}(n))$ of the current node. 
If the new point is dominated by $\widehat{z_*}(\mathcal{S}(n)$ it is immediately rejected (\textbf{Property 1}). If $\widehat{z^*}(\mathcal{S}(n)$ is covered, the node and its whole sub-tree is deleted (\textbf{Property 2}). Otherwise if $\widehat{z^*}(\mathcal{S}(n)) \succeq y$ or $y \succeq \widehat{z_*}(\mathcal{S}(n))$, the node needs to be analyzed. If $n$ is an internal node we call the algorithm recursively for each child. If $n$ is a leaf node, $y$ may be dominated by or dominate some points of $n$ and it is necessary to browse the whole list $\mathcal{L}(n)$ of the node $n$. If a point dominating $y$ is found, $y$ is rejected, and if a point dominated by $y$ is found, the point is deleted from $\mathcal{L}(n)$. 

If after checking ND-Tree the new point was found to be non-dominated it is inserted by adding it to a close leaf (Algorithm~\ref{algoInsert}). To find a proper leaf we start from the root and always select a child with closest distance to $y$. As a distance measure we use the Euclidean distance to the \emph{middle point}, i.e. a point lying in the middle of line segment connecting approximate ideal and approximate nadir points. 

Once we have reached a leaf node, we add the point $y$ to the list $\mathcal{L}(n)$ of the node and possibly update the ideal and nadir points of the node $n$ (Algorithm~\ref{algoUpdateIdealNadir}). However, if the size of $\mathcal{L}(n)$ became larger than the maximum allowed size of a leaf set, we need to split the node into a predefined number of children. To create children that contain points that are more similar to each other than to those in other children, we use a simple clustering heuristic based on Euclidean distance (see Algorithm~\ref{algoSplit}).

The approximate local ideal and nadir points are updated only when a point is added. We do not update them when point(s) are removed since it is a more complex operation. This is why we deal with approximate (not exact) local ideal and nadir points.

\floatname{algorithm}{Algorithm}
\begin{algorithm}[!ht]
\caption{\texttt{ND-Tree-based update}}\label{algoUpdate}
\begin{algorithmic}
\STATE Parameter $\updownarrow$: A Pareto archive $Y_N$ organized as ND-Tree 
\STATE Parameter $\downarrow$: New candidate point $y$ 

\vspace*{1\baselineskip}
\IF{$Y_N = \emptyset$}
\STATE Create a leaf node $n$ with $\mathcal{L}(n) = \{y\}$ and use it as a root
\ELSE 
\STATE $n \leftarrow$ root node
\IF{\texttt{UpdateNode}($n \updownarrow $,$y \downarrow$)}
\STATE \texttt{Insert}($n \updownarrow $,$y \downarrow$)
\ENDIF
\ENDIF
\end{algorithmic}
\end{algorithm}

\floatname{algorithm}{Algorithm}
\begin{algorithm}[!ht]
\caption{\texttt{UpdateNode}}\label{algoUpdateNode}
\begin{algorithmic}
\STATE Parameter $\updownarrow$: A node $n$ 
\STATE Parameter $\downarrow$: New candidate point $y$ 
\STATE Parameter $\uparrow$: Boolean (True if $y$ is not dominated by any points in the tree  of root $n$, False otherwise)
\vspace*{1\baselineskip}
\IF{$\widehat{z_*}(\mathcal{S}(n)) \succeq y$}
\STATE -$\,$-$|$ Property 1,  $y$ is rejected
\RETURN False
\ELSIF{$y \succeq \widehat{z^*}(\mathcal{S}(n))$}
\STATE -$\,$-$|$ Property 2
\STATE Remove $n$ and its whole sub-tree
\ELSIF{$\widehat{z^*}(\mathcal{S}(n)) \succeq y$ \OR $y \succeq \widehat{z_*}(\mathcal{S}(n))$}
\IF{$n$ is a leaf node}
\FORALL{$z \in \mathcal{L}(n)$} 
\IF{$z \succeq y$}
\RETURN False
\ELSIF{$y \succ z$}
\STATE $\mathcal{L}(n) \leftarrow \mathcal{L}(n) \backslash \{z\}$
\ENDIF
\ENDFOR
\ELSE
\FORALL{Child $n'$ of $n$} 
\IF{\NOT \texttt{UpdateNode} ($n' \updownarrow $,$y \downarrow$)}
\RETURN False
\ELSE
\IF{$n'$ became empty}
\STATE Remove $n'$ 
\ENDIF
\ENDIF
\ENDFOR
\IF{there is only one child $n'$ remaining}
\STATE Remove node $n$ and use $n'$ in place of $n$
\ENDIF
\ENDIF
\ELSE
\STATE -$\,$-$|$ Property 3
\STATE Skip this node
\ENDIF
\RETURN True
\end{algorithmic}
\end{algorithm}

\floatname{algorithm}{Algorithm}
\begin{algorithm}[!ht]
\caption{\texttt{Insert}}\label{algoInsert}
\begin{algorithmic}
\STATE Parameter $\updownarrow$: A node $n$ 
\STATE Parameter $\downarrow$: New candidate point $y$ 

\vspace*{1\baselineskip}
\IF{$n$ is a leaf node}
\STATE \texttt{$\mathcal{L}(n) \leftarrow \mathcal{L}(n) \cup \{y\}$}
\STATE \texttt{UpdateIdealNadir} ($n \updownarrow$,$y \downarrow$)
\IF{Size of $\mathcal{L}(n)$ became larger than maximum size of a leaf set}
\STATE \texttt{Split} ($n \updownarrow$)
\ENDIF
\ELSE
\STATE Find child $n'$ of $n$ being closest to $y$
\STATE \texttt{Insert}($n' \updownarrow $,$y \downarrow$)
\ENDIF
\end{algorithmic}
\end{algorithm}

\floatname{algorithm}{Algorithm}
\begin{algorithm}[!ht]
\caption{\texttt{Split}}\label{algoSplit}
\begin{algorithmic}
\STATE Parameter $\updownarrow$: A node $n$ 

\vspace*{1\baselineskip}
\STATE Find the point $z \in \mathcal{L}(n)$ with the highest average Euclidean distance to all other points in $\mathcal{L}(n)$
\STATE Create a new child $n'$ with list set $\mathcal{L}(n') = \{z\}$
\STATE $\mathcal{L}(n) \leftarrow \mathcal{L}(n) \backslash \{z\}$
\STATE \texttt{UpdateIdealNadir} ($n' \updownarrow$,$z \downarrow$) 
\WHILE{The required number of children are not created} 
\STATE{Find the point $z \in \mathcal{L}(n)$ with the highest average Euclidean distance to all points in all children of $n$} 
\STATE Create a new child $n'$ with an empty list set  $\mathcal{L}(n')$
\STATE $\mathcal{L}(n') \leftarrow \mathcal{L}(n') \cup \{z\}$
\STATE \texttt{UpdateIdealNadir} ($n' \updownarrow$,$z \downarrow$) 
\STATE $\mathcal{L}(n) \leftarrow \mathcal{L}(n) \backslash \{z\}$
\ENDWHILE
\WHILE{$\mathcal{L}(n)$ is not empty}
\STATE $z \leftarrow$ first point in $\mathcal{L}(n)$
\STATE Find child $n'$ of $n$ being closest to $z$
\STATE $\mathcal{L}(n') \leftarrow \mathcal{L}(n') \cup \{z\}$
\STATE \texttt{UpdateIdealNadir} ($n' \updownarrow$,$z \downarrow$) 
\STATE $\mathcal{L}(n) \leftarrow \mathcal{L}(n) \backslash \{z\}$
\ENDWHILE
\end{algorithmic}
\end{algorithm}

\floatname{algorithm}{Algorithm}
\begin{algorithm}[!ht]
\caption{\texttt{UpdateIdealNadir}}\label{algoUpdateIdealNadir}
\begin{algorithmic}
\STATE Parameter $\updownarrow$: A node $n$ 
\STATE Parameter $\downarrow$: New candidate point $y$ 

\vspace*{1\baselineskip}
\STATE Check in any component of $y$ is lower than corresponding component in $\widehat{z^*}(\mathcal{S}(n))$ or greater than corresponding component in $\widehat{z_*}(\mathcal{S}(n))$ and update the points if necessary
\IF {$\widehat{z^*}(\mathcal{S}(n))$ or $\widehat{z_*}(\mathcal{S}(n))$ have been changed}
\IF {$n$ is not a root}
\STATE $np \leftarrow$ parent of $n$
\STATE \texttt{UpdateIdealNadir} ($np \updownarrow$,$y \downarrow$)
\ENDIF
\ENDIF
\end{algorithmic}
\end{algorithm}

\subsection{Comparison to existing methods}

Like other methods ND-Tree-based update uses a tree structure to speed up the process of updating the Pareto archive. The tree and its use is, however, quite different from Quad-tree or k-d tree used in M-Front. For example, both Quad-tree or k-d tree partition the objective space based on comparisons on particular objectives, while in ND-Tree the space is partitioned based on the distances of points. Both Quad-tree or k-d tree have strictly defined degrees. In k-d tree it is always two (binary tree) while in Quad-tree it depends on the number of objectives. In ND-Tree the degree is a parameter. In Quad-tree the points are kept in both internal nodes and leaves, while ND-Tree keeps points in leaves only. In M-Front k-d tree is used to find an approximate nearest neighbor and the Pareto archive is updated using other structures (sorted lists for each objective). In our case, ND-Tree is the only data structure used by the algorithm.

\subsection{Computational complexity}

\subsubsection{Worst case}
\begin{figure}[!ht]
  \centering
  \includegraphics[scale=0.55,angle=0]{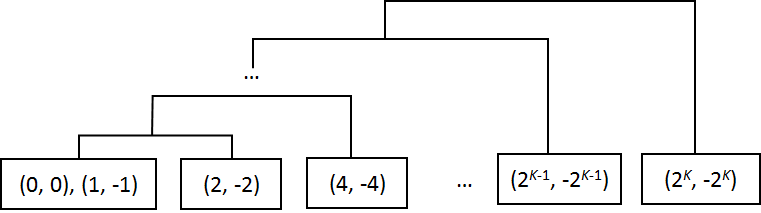}
   \caption{Example of fully unbalanced ND-Tree.}
   \label{fig:worst_case}
\end{figure}

The worst case for the \texttt{UpdateNode} and \texttt{Insert} algorithm is when we need to compare the new point to each intermediate node of the tree.
For example, consider the following particular case: two objectives, ND-Tree with maximum leaf size equal to 2, 2 children, and constructed by processing the following list of points (with $K \in \mathbb{N}^*$): (0, 0), (1, -1), ($2^K$, $-2^K$), ($2^{K-1}$, $-2^{K-1}$),..., (4, -4), (2, -2).
This list of points is constructed in such a way that the first two points are put in one leftmost leaf and the third point creates a separate leaf. Then each further point is closer to the child on the left side but finally after splitting the leftmost node the new point creates a new leaf. The ND-Tree obtained is shown in  Figure~\ref{fig:worst_case}.

Consider now that the archive is updated with point (0.5, -0.5). This point will need to be compared to all $N-1$ intermediate nodes and then to both points in leftmost leaf. In this case:

\begin{equation}
T(N) = 4 + T(N-1)
\end{equation}
where $N$ is the number of points in the archive and $T(N)$ is the number of point comparisons needed to update an archive of size $N$. Term $4$ appears because we check two children and for each child approximate ideal and nadir points are compared. Solving the recurrence we get:
\begin{equation}
T(N) = 4 + 4N
\end{equation}
Thus, the algorithm has $\mathcal{O}(N)$ time complexity.

We are not aware of any result showing that the worst-case time complexity of any algorithm for the dynamic non-dominance problem may be lower than $\mathcal{O}(N)$ in terms of point comparisons. So, our method does not improve the worst-case complexity but according to our experiments performs significantly better in practical cases.

\subsubsection{Best case}

Assume first that the candidate point is not covered by any point in $Y_N$. In the optimistic case, at each intermediate node the points are equally split into predefined number of children and there is only one child that has to be further processed (i.e. there is only one child for which none of the three properties hold). In fact we could consider even more optimistic distribution of points when the only node that has to be processed contains just one point, but the equal split is much more realistic assumption. In this case:
\begin{equation}
T(N) = 2C + T(N/C) = \Theta(\log_C N)
\end{equation}
where $C$ is the number of children. If the candidate point is covered by a point in $Y_N$ the \texttt{UpdateNode} algorithm may stop even earlier and there will be no need to run \texttt{Insert} algorithm.

\subsubsection{Average case}

Analysis and even definition of average case for such complex algorithms is quite difficult. The simplest case to analyze is when each intermediate node has two children which allows us to follow the analysis of well-known algorithms like binary search or Quicksort. If a node has $N$ points then one of the two children may have $1,\dots,N-1$ points and the other child the remaining number of points. Assuming that each split has equal probability and only one child is selected:
\begin{equation} \label{eq:a1}
T(N)= 4 + \frac{1}{N} \sum_{k=1}^{N-1}T(k) \\
\end{equation}
Multiplying both sides by $N$:
\begin{equation} \label{eq:a2}
NT(N)= 4N + \sum_{k=1}^{N-1}T(k) \\
\end{equation}
Assuming that $N \geq 2$:
\begin{equation} \label{eq:a3}
(N-1)T(N-1) = 4(N-1) + \sum_{k=1}^{N-2}T(k)
\end{equation}
Subtracting equations \ref{eq:a2} and \ref{eq:a3}:
\begin{equation} \label{eq:a4}
\begin{split}
NT(N) - (N-1)T(N-1) = \\
4N - 4(N-1) + \sum_{k=1}^{N-1}T(k) - \sum_{k=1}^{N-2}T(k) 
\end{split}
\end{equation}
Simplifying:
\begin{equation} \label{eq:a5}
T(N) = \frac{4}{N} + T(N-1)
\end{equation}
Solving this recurrence:
\begin{equation} \label{eq:a6}
T(N) = 4H_N-1
\end{equation}
where $H_N$ is N-th harmonic number. Using well-known properties of harmonic numbers we get $T(N) = \Theta(\log N)$.

We can expect, however, that in realistic cases more than one child will need to be further processed in \texttt{UpdateNode} algorithm because the candidate point may cover approximate nadir points or may be covered by approximate ideal points of more than one child. Assume that the probability of selecting both children is equal to $c_1$. Then:
\begin{equation} \label{eq:c1}
T(N)= 4 + \frac{1}{N} (1 + c_1)\sum_{k=1}^{N-1}T(k) \\
\end{equation}
Following the above reasoning we get:
\begin{equation} \label{eq:c2}
T(N) = \frac{2}{N} + \frac{N+c_1}{N}T(N-1)
\end{equation}
Solving this recurrence, we obtain:
\begin{equation} \label{eq:c3}
\begin{split}
T(N) = \frac{\Gamma(c_1+N+1)}{\Gamma(N+1)} - \frac{2}{c_1}
\end{split}
\end{equation}
(that can be checked by substituting (\ref{eq:c3}) in (\ref{eq:c2})). 
	
\noindent Since
\begin{equation} \label{eq:c4}
\lim_{N\to \infty} \frac{\Gamma(N+\alpha)}{\Gamma(N)N^\alpha} = 1
\end{equation}
We have:
\begin{equation} \label{eq:c5}
T(N)= \Theta(N^{c_1})
\end{equation}
and the algorithm remains sub-linear for any $c_1 < 1$. In the worst case, both children need to be selected, so $c_1 = 1$ and T$(N)= \Theta(N)$ which confirms the analysis presented above.

This analysis may give only approximate insight into behaviour of the algorithm since in \texttt{UpdateNode} algorithm $c_1$ will not be constant at each level. We may rather expect that while going down the tree from the root towards leaves the probability that two children will need to be processed will decrease because approximate nadir and ideal points of the children will lie closer. Anyway, this analysis shows that the performance of \texttt{UpdateNode} algorithm may be improved by decreasing the probability that a child has to be processed. This is why we try to locate in one node points lying close to each other in \texttt{Insert} and \texttt{Split} algorithms.

In \texttt{Insert} algorithm always one child is processed, so the time complexity remains $\Theta(\log N)$ in average case.

The main part of \texttt{Split} algorithm has constant time complexity since it depends only on the maximum size of
a leaf set which is a constant parameter of the algorithm.

\texttt{UpdateIdealNadir} algorithm goes up the tree starting from a leaf which is equivalent to going down the tree and selecting just one child. So, its analysis is exactly the same as of \texttt{Insert} algorithm.

We also need to consider the complexity of the operation of removal of node $n$ and its sub-tree. In the worst case, the removed node is the root, and thus all $N$ point need to be removed. Such situation is very unlikely, since it happens when the new point dominates all points in the current archive. Typically, the new point will dominate only few points. 

\section{Computational experiments}

We will show results obtained with ND-Tree and other methods in two different cases:
\begin{enumerate}[label=\Alph*)]
\item Results for artificially generated sets which allow us to easily control the number of points in the sets and the quality of the points.
\item Results for sets generated by a MOEA, namely MOEA/D~\cite{zhang2007moea} for the multiobjective knapsack problem.
\end{enumerate}

We compare the simple list, sorted list (biobjective case), Quad-tree, M-Front and ND-Tree for these sets according to the CPU time [ms]. To avoid the influence of implementation details all methods were implemented from the scratch in C++ in as much homogeneous way as possible, i.e. when possible the same code was used to perform the same operations like Pareto dominance checks.  

For the implementation of Quad-tree, we use Quad-tree2 version as described by Mostaghim and Teich~\cite{Mostaghim2004}.

For M-Front, we use as much as possible the description found in the paper of Drozd\'ik \emph{et al.}~\cite{Drozdik2015}. However the authors do not give the precise algorithm of k-d tree used in their method. In our implementation of k-d tree, when a leaf is reached a new division is made using the average value of the current level objective. The split value is average between the value of new point and the point in the leaf. Also like in Drozd\'ik \emph{et al.}~\cite{Drozdik2015} the approximate nearest neighbor is found exactly as in the standard exact nearest neighbor search, but only four evaluations of the distance are allowed. Note that at \url{https://sites.google.com/site/ndtreebasedupdate/} we present results of an additional experiment showing that for a higher number of objectives the details of the implementation of k-d tree do not have any substantial influence on the running time. 

We also noticed that a number of elements of M-Front can be improved to further reduce the running time. The improvement is in our opinion significant enough to call the new method M-Front-II. In particular in some cases M-Front-II was several times faster than original M-Front in our computational experiments. The modifications we introduced are as follows:
\begin{itemize}
\item In original M-Front the sets $RS_L(y, ref)$ and $RS_U(y, ref)$ are built explicitly and only then the points contained in these sets are compared to $y$. In M-Front-II we build them only implicitly, i.e. we immediately compare the points that would be added to the sets to $y$.
\item We analyze the objectives in such a way that we start with objectives for which 
$ref_k \leq y_k$. In other words, we start with points from set $RS_U(y, ref)$. Since many new points are dominated this allows to stop the search immediately when a point dominating $y$ is found. Note that a similar mechanism is in fact used in original M-Front but only after the sets 
$RS_L(y, ref)$ and $RS_U(y, ref)$ are explicitly built.
\item The last modification is more technical. M-Front uses linked lists (\verb'std::list' in C++) to store the indexes and a hash-table (\verb'std::unordered_map' in C++) to link points with their positions in these lists. We observed, however, that even basic operations like iterating over the list are much slower with linked lists than with static or dynamic arrays (like \verb'std::vector' in C++). Thus we use dynamic arrays for the indexes. In this case, however, there is no constant iterator that could be used to link the points with their positions in these indexes. So, we use a binary search to locate a position of a point in the sorted index whenever it is necessary. The overhead of the binary search is anyway smaller than the savings due to the use of faster indexes.
\end{itemize}

For ND-Tree we use 20 as the maximum size of a leaf and $p+1$ as the number of children. These values of the parameters were found to perform well in many cases. We analyze the influence of these parameters later in this section.

The code, as well as test instances and data sets, are available at \url{https://sites.google.com/site/ndtreebasedupdate/}. All of the results have been obtained on an Intel Core i7-5500U CPU at 2.4 GHz. 
\vspace{-0.5cm}
\subsection{Artificial sets}

\subsubsection{Basic, globally convex sets}

The artificial sets are composed of $n$ points with $p$ objectives to minimize. The sets are created as follows. We generate randomly $n$ points $y^i$ in $\{0,\ldots,V_{max}\}^p$ with the following constraint: 
$\sum_{k=1}^p (V_{max}-y^i_k)^2 \leq V_{max}^2$. With this constraint, all the non-dominated points will be located inside the hypersphere with the center at $(V_{max},\ldots,V_{max})$ and with a radius of length equal to $V_{max}$. In order to control the quality of the generated points, we also add a quality constraint:
$\sum_{i=k}^p (V_{max}-y^i_k)^2 \geq (1-\epsilon)*V_{max}^2$. In this way, with a small $\epsilon$, only high-quality points will be generated. We believe that it is a good model for points generated by real MOEAs since a good MOEA should generate points lying close to the true Pareto front. The hypersphere is a model of the true Pareto front and parameter $\epsilon$ controls the maximum  distance from the hypersphere. We have generated data sets composed of 100 000 and 200 000 points, with $V_{max}=10000$, and for $p=2$ to 10. In the main experiment we use data sets with 100 000 points because for the larger sets running times of some methods became very long. Also because of very high running times for sets with many objectives, in the main experiment we used sets with up to 6 objectives. For each value of $p$, five different quality levels are considered: quality q1, $\epsilon=0.5$ ; q2, $\epsilon=0.25$ ; q3, $\epsilon=0.1$ ; q4, $\epsilon=0.05$ ; q5, $\epsilon=0.01$. The fraction of non-dominated points grows both with increasing quality and number of objectives and in extreme cases all points may be non-dominated (see Table~\ref{tab:numbersnd}).

\begin{table}[!ht]
\caption{Numbers of non-dominated points in artificial sets}
\begin{center}
\label{tab:numbersnd}
\begin{tabular}{cccccc}
$p$ & Quality & $|Y_N|$ & $|Y_N|$ & $|Y_N|$ \\ 
& & convex & non-convex &  clustered \\ 
\hline
2	&	q1	&	519	&	379	&	449	\\
2	&	q2	&	713	&	613	&	552	\\
2	&	q3	&	1046	&	1037	&	785	\\
2	&	q4	&	1400	&	1454	&	1059	\\
2	&	q5	&	2735	&	2748	&	1781	\\
3	&	q1	&	4588	&	2587	&	3729	\\
3	&	q2	&	6894	&	5344	&	5514	\\
3	&	q3	&	12230	&	11497	&	9720	\\
3	&	q4	&	19095	&	18648	&	15502	\\
3	&	q5	&	53813	&	53255	&	44173	\\
4	&	q1	&	14360	&	6853	&	11963	\\
4	&	q2	&	21680	&	16420	&	18120	\\
4	&	q3	&	39952	&	37709	&	35460	\\
4	&	q4	&	64664	&	63140	&	57725	\\
4	&	q5	&	98283	&	98243	&	97137	\\
5	&	q1	&	28944	&	13437	&	23966	\\
5	&	q2	&	42246	&	34357	&	38028	\\
5	&	q3	&	77477	&	75796	&	71063	\\
5	&	q4	&	96002	&	95867	&	93842	\\
5	&	q5	&	99975	&	99975	&	98521	\\
6	&	q1	&	45879	&	22956	&	40483	\\
6	&	q2	&	65195	&	57966	&	61096	\\
6	&	q3	&	96687	&	96480	&	94978	\\
6	&	q4	&	99788	&	99786	&	99652	\\
6	&	q5	&	100000	&	100000	&	99999	\\
\end{tabular}
\end{center}
\end{table}

\subsubsection{Globally non-convex sets}

In order to test whether the global convexity of the above sets influences the behavior of the tested methods we have also generated sets whose Pareto fronts are globally non-convex. They were obtained by simply changing the sign of each objective in the basic sets.

\subsubsection{Clustered sets}

In these sets, the points are located in small clusters. We have generated sets composed of 100 clusters, where each cluster contains 1000 points (the sets are thus composed of 100 000 points). The sets have been obtained as follows: we start with the 200 000 points from the basic convex sets. We select from the set one random point, and we then select the 999 points closest to this point (using the Euclidean distance). We repeat this operation 100 times to obtain the 100 clusters of the sets.   

The shapes of exemplary biobjective globally convex, globally non-convex and clustered data sets can be seen at \url{https://sites.google.com/site/ndtreebasedupdate/}. 

\begin{figure}[!hbp]
  \centering
  \includegraphics[scale=0.55,angle=0]{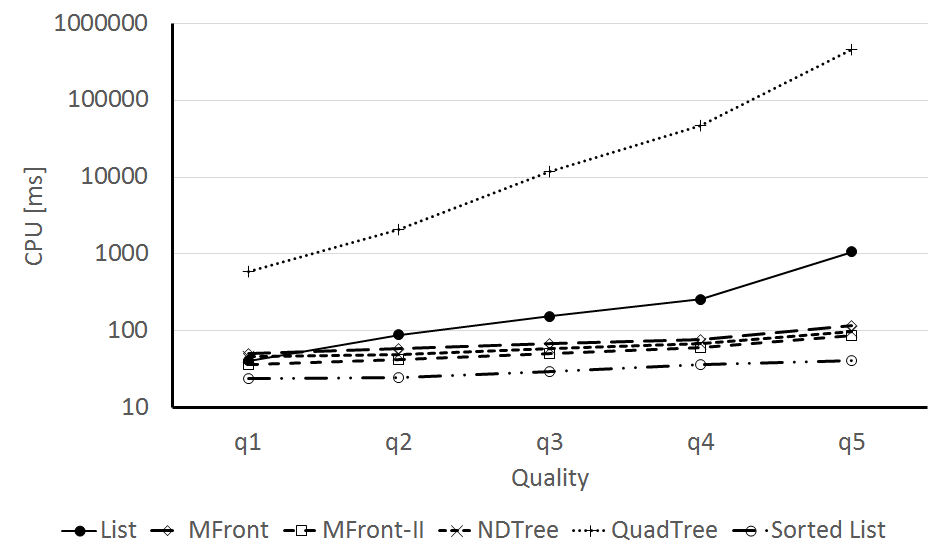}
   \caption{CPU time (logarithmic scale) for 
   biobjective convex data sets.}
   \label{fig:obj2}
\end{figure}

\begin{figure}[!hbp]
  \centering
  \includegraphics[scale=0.55,angle=0]{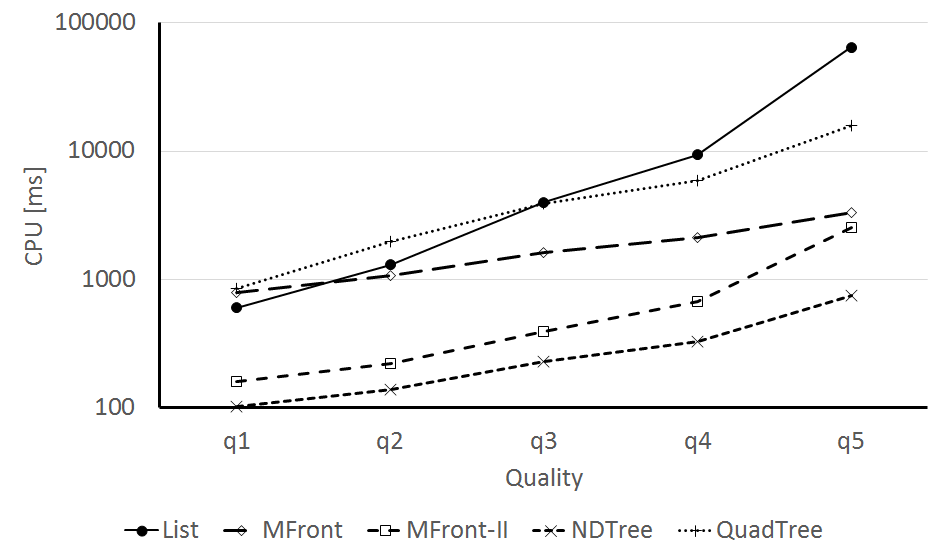}
   \caption{CPU time (logarithmic scale) for three-objective convex data sets.}
   \label{fig:obj3}
\end{figure}

\begin{figure}[!hbp]
  \centering
  \includegraphics[scale=0.55,angle=0]{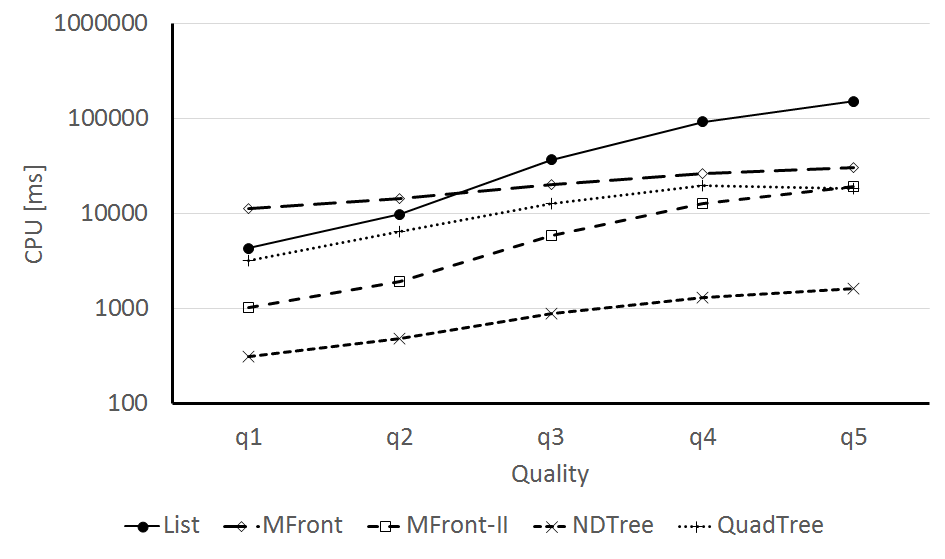}
   \caption{CPU time (logarithmic scale) for four-objective convex data sets.}
   \label{fig:obj4}
\end{figure}

\begin{figure}[!hbp]
  \centering
  \includegraphics[scale=0.55,angle=0]{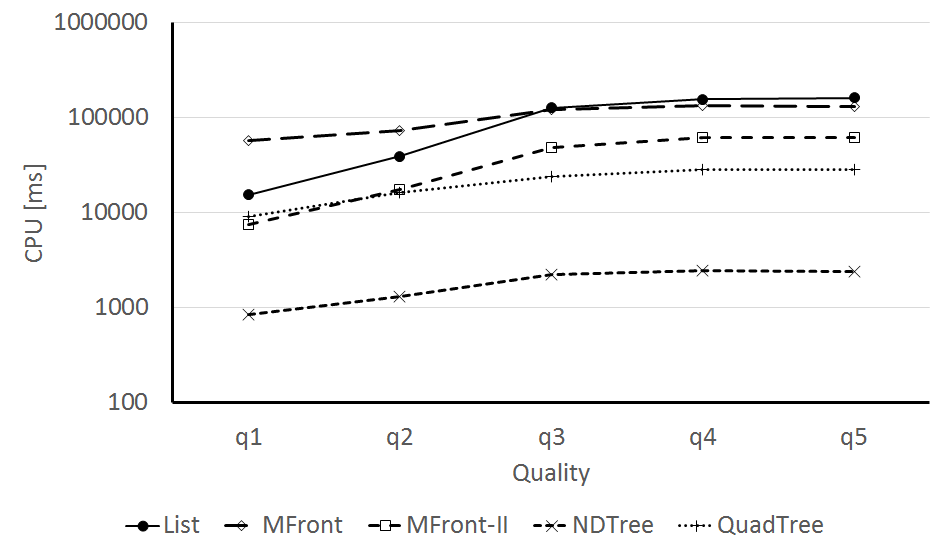}
   \caption{CPU time (logarithmic scale) for five-objective convex data sets.}
   \label{fig:obj5}
\end{figure}

\begin{figure}[!hbp]
  \centering
  \includegraphics[scale=0.55,angle=0]{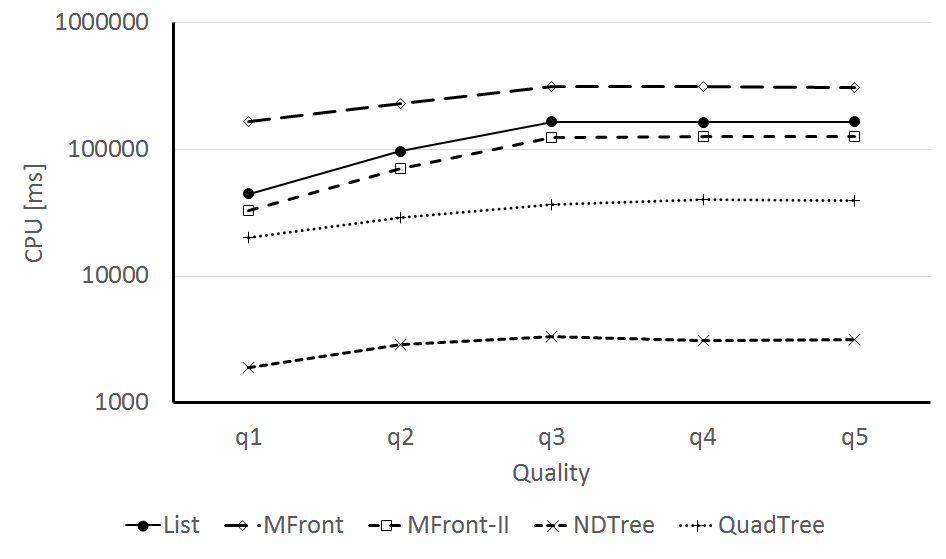}
   \caption{CPU time (logarithmic scale) for six-objective convex data sets.}
   \label{fig:obj6}
\end{figure}

Each method was run 10 times for each set, with the points processed in a different random order for each run. The average running times for basic sets are presented in Figures~\ref{fig:obj2} to~\ref{fig:obj6}. We use average values since the different values were generally well distributed around the average with small deviations. Please note that because of large differences the running time is presented in logarithmic scale. 

\begin{figure}[!hbp]
  \centering
  \includegraphics[scale=0.55,angle=0]{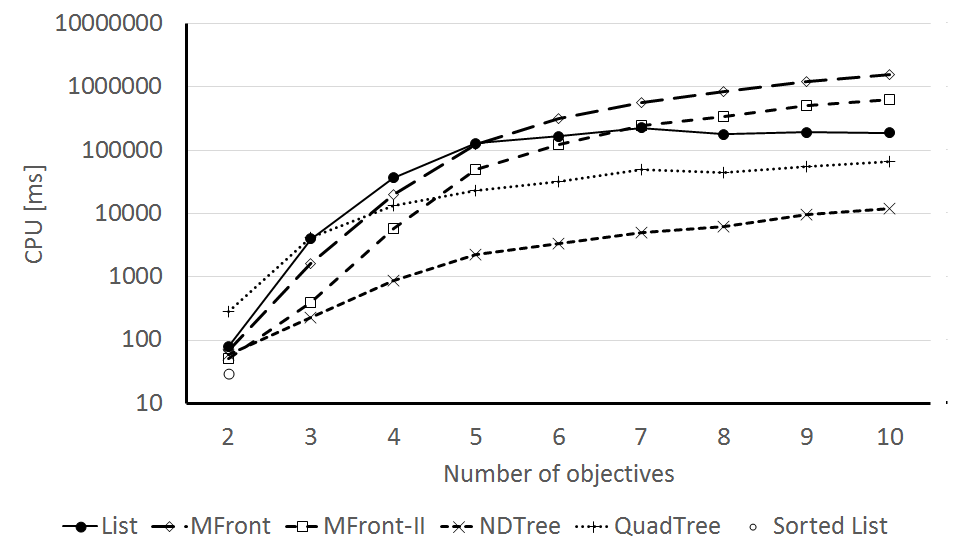}
   \caption{CPU time (logarithmic scale) vs. number of objectives for convex data sets of quality q3.}
   \label{fig:numobj}
\end{figure}

In addition, in Figure~\ref{fig:numobj} we illustrate the evolution of the running times according to the number of objectives for the data sets of intermediate quality q3. In this case we use sets with up to 10 objectives. With 7 and more objectives even the sets of intermediate quality q3 contain almost only non-dominated points (see Table~\ref{tab:numbersnd}). This is why the running times of the simple list are practically constant for 7 and more objectives, because the simple list boils down in this case to the comparison of each new point to all points in the list. The running times of all other methods including ND-Tree increase with a growing number of objectives, but ND-Tree remains 5.5 times faster than the second best method (Quad-tree) for sets with 10 objectives.

\begin{table}[!hbp]
\caption{Point comparisons per ms}
\begin{center}
\label{tab:FPC_ms}
\begin{tabular}{ccc}
Method & Comparisons per ms \\ \hline
List	&	26 752 \\
M-Front	&	3 476 	\\
M-Front-II	&	8 370 	\\
ND-Tree	&	17 733	\\
Quad-tree	&	9 040	
\end{tabular}
\end{center}
\end{table}

\begin{figure}[!hbp]
  \centering
  \includegraphics[scale=0.55,angle=0]{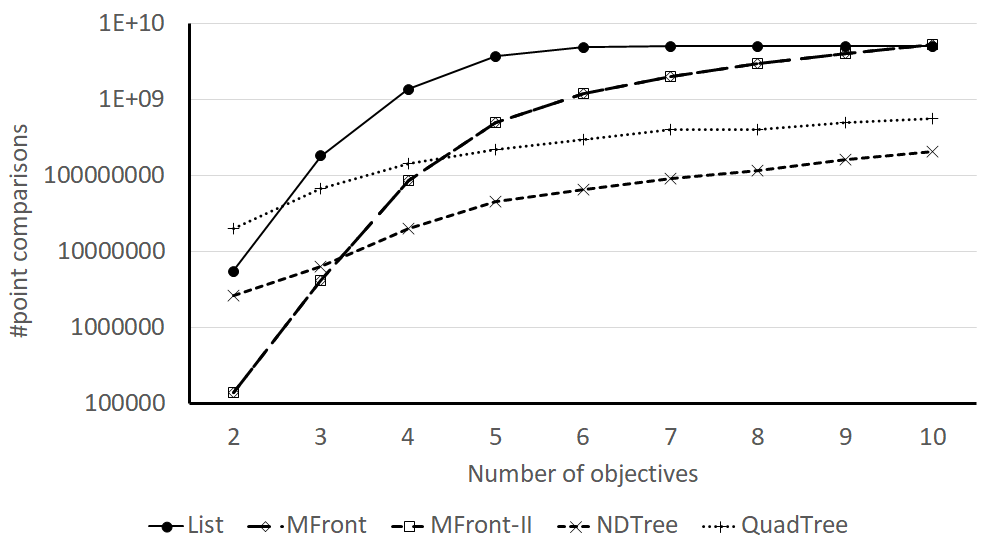}
   \caption{Number of point comparisons (logarithmic scale) vs. number of objectives for convex data sets of quality q3.}
   \label{fig:ops}
\end{figure}

Furthermore, we measured the number of comparisons of points with the dominance relation for the data sets of intermediate quality q3 with $p=2,...,10$. For ND-Tree-based update it includes also comparisons to the approximate ideal and nadir points and for Quad-tree comparisons to sub-nodes. The results are presented in Figure \ref{fig:ops}. Please note that the results for the two versions of M-Front overlap in the Figure.

The differences in running times cannot be fully explained by the number of point comparisons because the methods differ significantly in the number of point comparisons per milliseconds (see Table \ref{tab:FPC_ms}). This ratio is highest for the simple list because this method performs very few additional operations. It is also relatively high for ND-Tree. Other methods perform many other operations than point comparisons that strongly influence their running times. This is particularly clear in comparison of M-Front and M-Front-II. These method perform the same number of point comparisons, but M-Front-II is several times faster than M-Front because in the latter method the sets $RS_L(y, ref)$ and $RS_U(y, ref)$ are built explicitly and this method uses slower linked lists. Overall, ND-Tree performs fewest number of point comparisons for data sets with $p \geq 4$. These results indicate that ND-Tree-based update substantially reduces the number of comparisons with respect to the simple list. For example for $p=10$, all points in the data set are non-dominated, thus on average each of the $100 000$ new points has to be compared to an archive composed of $49999.5$ points, while with ND-Tree-based update it only requires $2029$ point comparisons on average. 

\begin{figure}[!hbp]
  \centering
  \includegraphics[scale=0.55,angle=0]{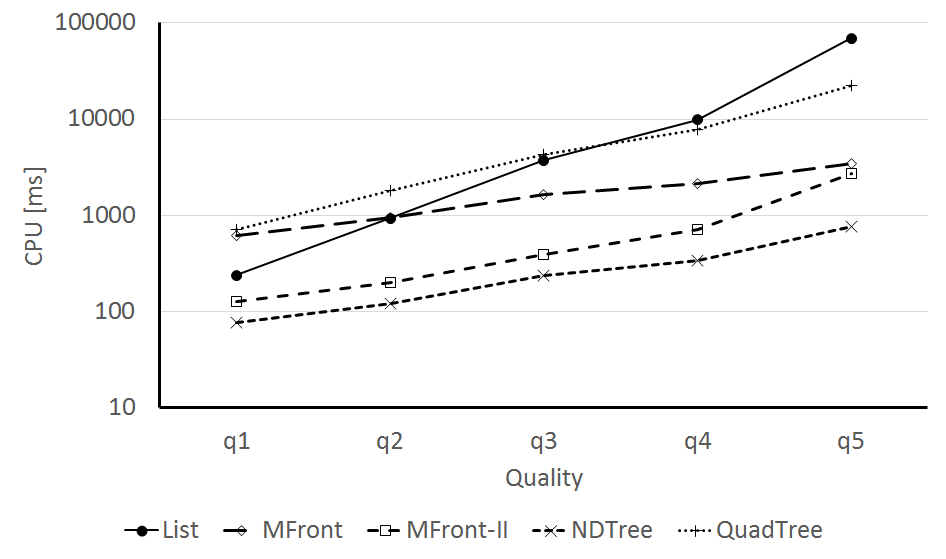}
   \caption{CPU time (logarithmic scale) for three-objective non-convex data sets.}
   \label{fig:obj3-non-convex}
\end{figure}

\begin{figure}[!hbp]
  \centering
  \includegraphics[scale=0.55,angle=0]{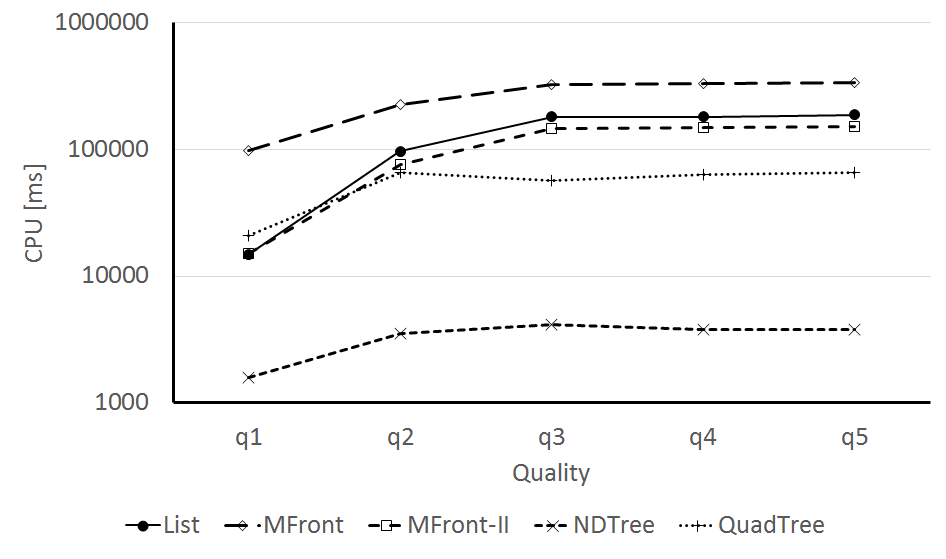}
   \caption{CPU time (logarithmic scale) for six-objective non-convex data sets.}
   \label{fig:obj6-non-convex}
\end{figure}

\begin{figure}[!hbp]
  \centering
  \includegraphics[scale=0.55,angle=0]{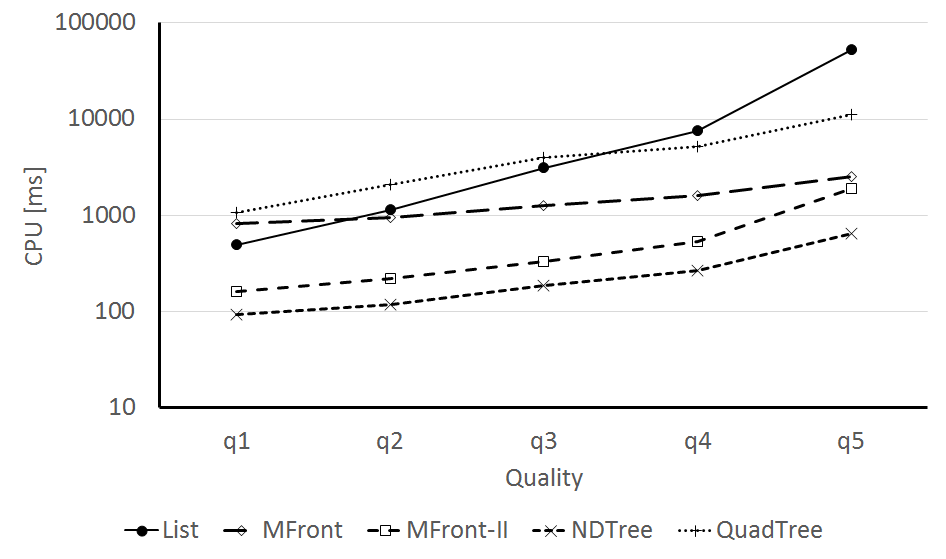}
   \caption{CPU time (logarithmic scale) for three-objective clustered data sets.}
   \label{fig:obj3-clustered}
\end{figure}

\begin{figure}[h]
  \centering
  \includegraphics[scale=0.55,angle=0]{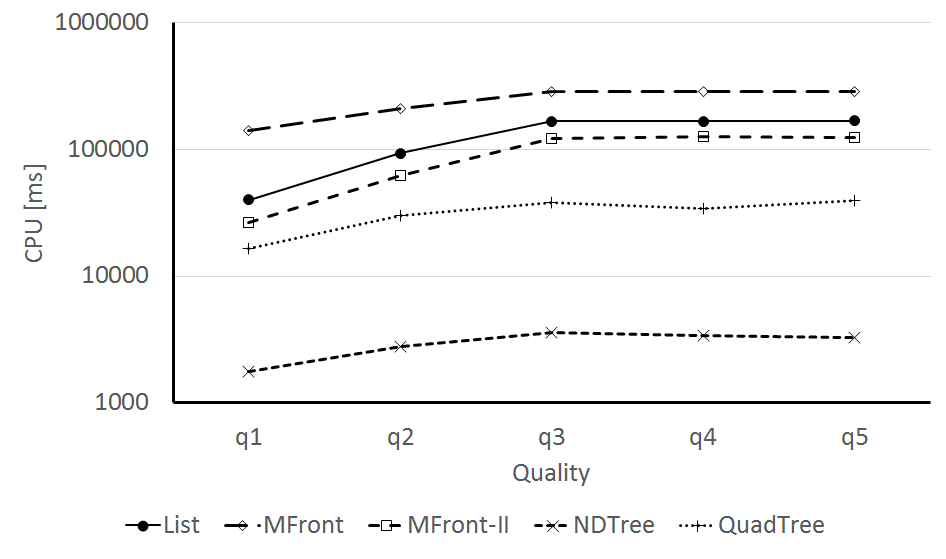}
   \caption{CPU time (logarithmic scale) for six-objective clustered data sets.}
   \label{fig:obj6-clustered}
\end{figure}

The results obtained for non-convex and clustered sets were very similar to the results with basic sets. Thus, in Figures~\ref{fig:obj3-non-convex} to~\ref{fig:obj6-clustered} we show only exemplary results for the three- and six-objective cases. These results indicate that the running times of the tested methods are not substantially affected by the global shape of the Pareto front.

\subsubsection{Discussion of the results for artificial sets}

The main observations from this experiment are:
\begin{itemize}
\item ND-Tree performs the best in terms of CPU time for all test sets with three and more objectives. In some cases the differences to other methods are of two orders of magnitude and in some cases the difference to the second best method is of one order of magnitude. ND-Tree behaves also very predictably, its running time grows slowly with increasing number of objectives and increasing fraction of non-dominated points.
\item For biobjective instances sorted list is the best choice. In this case, M-Front and M-Front-II also behave very well since they become very similar to sorted list.
\item The simple list obtains its best performances for data sets with many dominated points like $p=2$ with lowest quality. In this case the new point is often dominated by many points, so the search process is quickly stopped after finding a dominating point.
\item Quad-tree performs very badly for data sets with many dominated points, e.g. in biobjective instances where it is the worst method in all cases. In this case, many points added to Quad-tree are then removed and the removal of a point from Quad-tree is a costly operation. As discussed above when an existing point is removed its whole sub-tree has to be re-inserted to the structure. On the other hand, it is the second best method for most data sets with six and more objectives.
\item M-Front-II is much faster than M-Front on data sets with larger fraction of dominated points. In this case, M-Front-II may find a dominating point faster without building explicitly the whole sets $RS_L(y, ref)$ and $RS_U(y, ref)$.
\item The performance of both M-Front and M-Front-II deteriorates with an increasing number of objectives. With six and more objectives M-Front is the slowest method in all cases. Intuitively this can be explained by the fact that M-Front (both versions) uses each objective individually to reduce the search space. In the case of two objectives the values of one objective carry a lot of information since the order on one objective induces also the order on the other one. The more objectives, the less information we get from an order on one of them. Furthermore, sets $RS_L(y, ref)$ and $RS_U(y, ref)$ are in fact unions of corresponding sets for particular objectives, which also results in their growth. Finally, in many objective case, a reference point close on Euclidean distance does not need to be very close on each objective, since it will rather have a good balance of differences on many coordinates. 


\end{itemize}

\begin{figure}[!hbp]
  \centering
  \includegraphics[scale=0.55,angle=0]{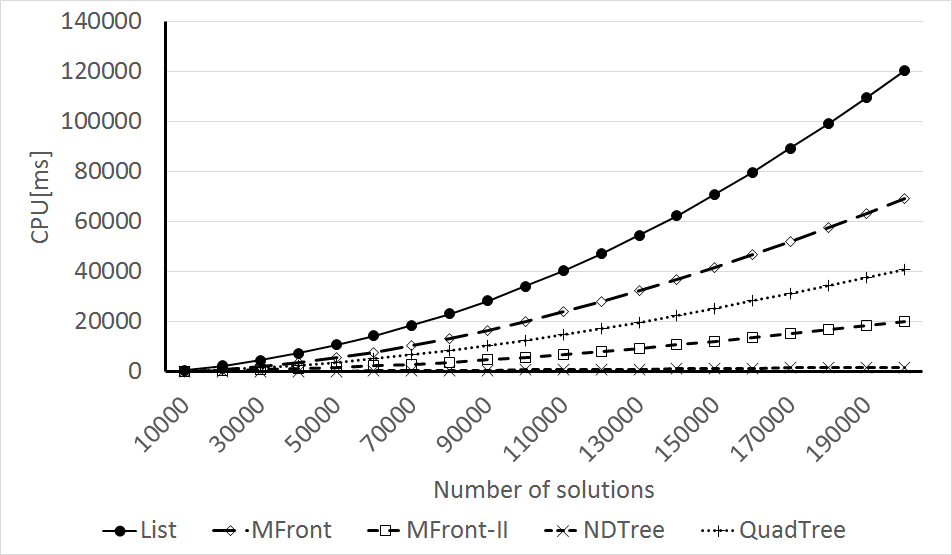}
   \caption{CPU time (linear scale) vs. the number of points for four-objective convex data sets of quality q3.}
   \label{fig:numberAll}
\end{figure}

\begin{figure}[!hbp]
  \centering
  \includegraphics[scale=0.55,angle=0]{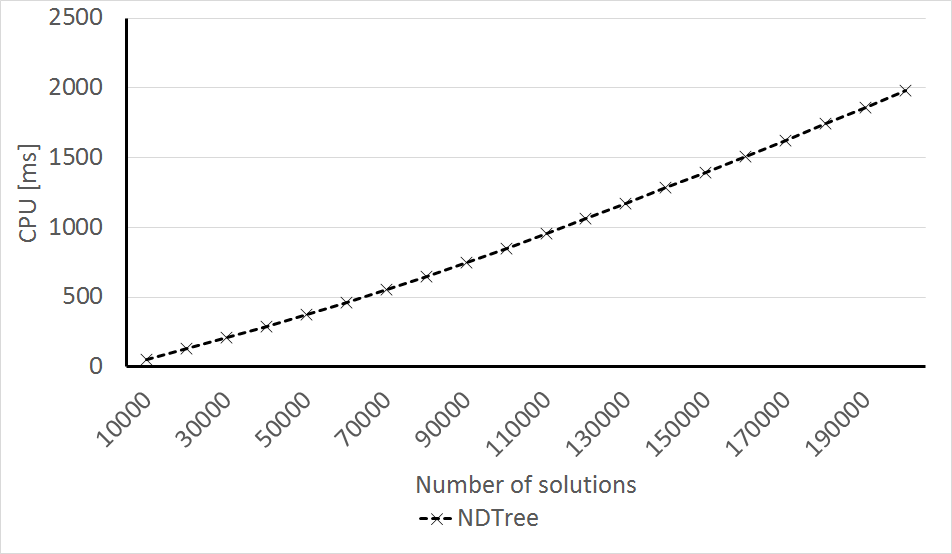}
   \caption{CPU time (linear scale) vs. the number of points for ND-Tree only (four-objective convex data sets of quality q3).}
   \label{fig:numberNDTree}
\end{figure}

In an additional experiment we analyzed the evolution of the running times of all methods with increasing number of points. We decided to use one intermediate globally convex data set with $p=4$ and quality q3. We used 200 000 points in this case and 10 runs for each method (see Figure~\ref{fig:numberAll}). The CPU time is cumulative time of processing a given number of points. In addition, since the running time is much smaller for ND-Tree its results are presented in Figure~\ref{fig:numberNDTree} separately. We see that ND-Tree is the fastest method for any number of points and its cumulative running time  grows almost linearly with the number of points. In other words, time of processing a single point is almost constant. Please note that unlike in other figures the linear scale is used in these two figures in order to make them more informative. 

\begin{figure}
  \centering
  \includegraphics[scale=0.55,angle=0]{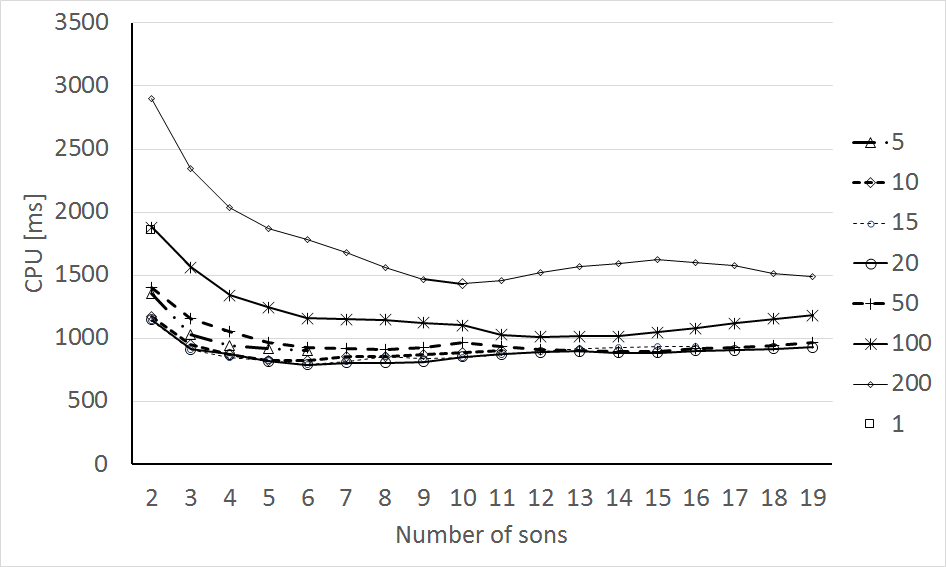}
   \caption{ND-Tree CPU time for different values of parameters (four-objective convex data sets of quality q3). Data series correspond to maximum leaf size. CPU time in linear scale.}
   \label{fig:parameters}
\end{figure}

ND-Tree has two parameters - the maximum size (number of points) of a leaf, and the number of children, so the question arises how sensitive it is to the setting of these parameters. To study it we again use the intermediate data set with $p=4$ and quality q3 and run ND-Tree with various parameters, see Figure~\ref{fig:parameters}. Please note that number of children cannot be larger than the maximum size of a leaf $+1$ since after exceeding the maximum size the leaf is split into the given number of children. We see that ND-Tree is not very sensitive to the two parameters: the CPU time remains between about one and three seconds regardless of the values of the parameters. The best results are obtained with 20 for the maximum size of a leaf and with 6 for the number of children. 

In our opinion the results confirm that ND-Tree performs relatively well for a wide range of the values of the parameters. In fact, it would remain the best method for this data set with any of the parameters settings tested.

\subsection{Sets generated by MOEA/D}

In order to test if the observations made for artificial sets hold for sets generated by real evolutionary algorithms, we use sets of points generated by well-known MOEA/D algorithm \cite{zhang2007moea} for multiobjective multidimensional knapsack problem instances with 2 to 6 objectives. We used the code available at \url{http://dces.essex.ac.uk/staff/zhang/webofmoead.htm} \cite{zhang2007moea}. We used the instances with 500 items available with the code with 2 to 4 objectives. The profits and weights of these instances were randomly generated uniformly between 1 and 100. We have generated ourselves the 5 and 6 objectives instances by adding random profits and weights, between 1 and 100. MOEA/D was run for at least 100 000 iterations and the first 100 000 points generated by the algorithm were stored for the purpose of this experiment. The numbers of non-dominated points are given in Table~\ref{tab:numbersMOEAD}. 

\begin{figure}
  \centering
  \includegraphics[scale=0.55,angle=0]{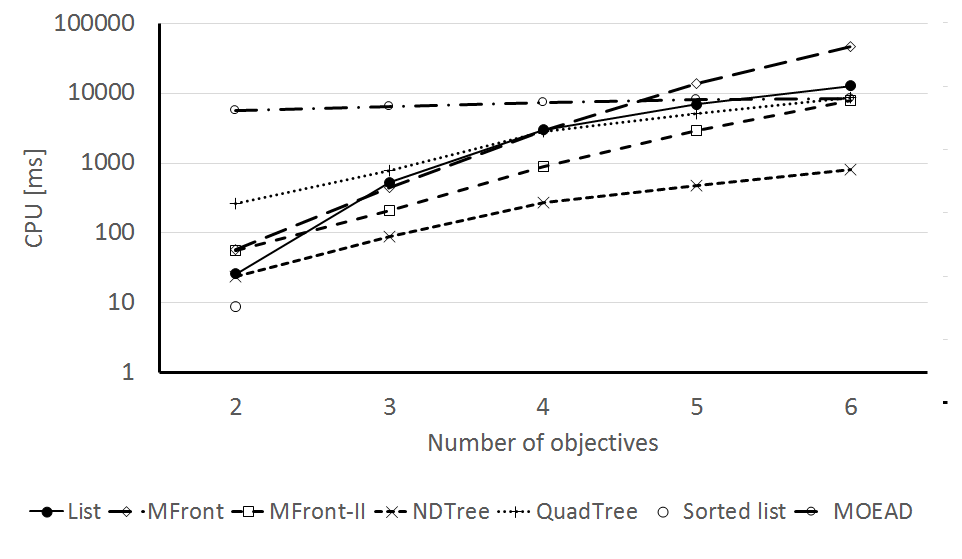}
   \caption{CPU time (logarithmic scale) on MOEA/D sets.}
   \label{fig:MOEAD}
\end{figure}

Figure~\ref{fig:MOEAD} presents running times of each of the tested methods as well the running times of MOEA/D excluding the time needed to update the Pareto archive. These results confirm that the observations made for artificial sets also hold in the case of real sets. ND-Tree is the fastest method for three and more objectives. Quad-tree performs particularly badly in the biobjective case. Both versions of M-Front relatively deteriorate with a growing number of objectives. Furthermore, these results show that the time of updating the Pareto archive may be higher than the remaining running time of MOEA/D. In particular, for the six-objective instance the running time of M-Front is 5 times higher than the remaining running time of MOEA/D. The running time of the simple list, Quad-tree and M-Front-II are comparable to the remaining running time of MOEA/D, and only the running time of ND-Tree is 10 times shorter. This confirms that the selection of an appropriate method for updating the Pareto archive may have a crucial influence on the running time of a MOEA. 

\begin{table}[t]
\caption{Numbers of non-dominated points in sets generated by MOEA/D}
\begin{center}
\label{tab:numbersMOEAD}
\begin{tabular}{ccc}
$p$ & $|Y_N|$ \\ \hline
2	&	140 \\
3	&	1789 	\\
4	&	5405 	\\
5	&	10126 	\\
6	&	16074 	
\end{tabular}
\end{center}
\end{table}

We have also generated sets of points by applying another MOMH, namely Pareto local search~\cite{Paquete04} to solve the multiobjective traveling salesman problem (MOTSP). These results can be found at \url{https://sites.google.com/site/ndtreebasedupdate/} (the same conclusions apply).

\section{ND-Tree-based non-dominated sorting}

ND-Tree-based update may also be applied to the problem of the non-dominated sorting. This problem arises in the non-dominated sorting-based MOEAs, e.g. NSGA-III \cite{Deb2014} where a population of solutions needs to be assigned to different fronts based on their dominance relationships.

We solve the non-dominated sorting problem in the very straightforward way, i.e. we find the first front by updating an initially empty Pareto archive with each point in the population. Then the points from the first front are removed from the population and the next front is found in the same way. This process is repeated until all fronts are found.

We compare this approach to some recently proposed non-dominated sorting algorithms, i.e. ENS-BS/SS \cite{Zhang2015} and DDA-NS \cite{Zhou2017}. ENS-BS/SS algorithm sorts the points lexicographically based on the values of objectives. Then it considers each solution using this order to efficiently find the last front that contains a point dominating the considered point. For each front this method in fact solves the dynamic non-dominance problem with the simple list and if the considered solution is non-dominated within this front it needs to be compared to each solution in this front. If there is just one front in the population, the method boils down to solving the dynamic non-dominance problem with the simple list and requires $\mathcal{O}(N^2)$ comparisons. DDA-NS algorithm sorts the population according to each objective which requires $\mathcal{O}(N\log N)$ objective function comparisons and builds comparison matrices for each objectives. Then it uses some matrix operations which in general have $\mathcal{O}(N^2)$ complexity to build the fronts. We also compare our algorithm to M-Front-II \cite{Drozdik2015} applied in the same way as ND-Tree-based update. Please note that similarly to what was done in \cite{Zhou2017} we apply M-Front-II for finding each front, while in \cite{Drozdik2015} only the first front was found by M-Front.

Also, like \cite{Zhang2015, Zhou2017} we used populations of size 5000. We used both random populations drawn with uniform probability from a hypercube, and populations composed of 5000 randomly selected points from our data sets of intermediate quality q3. For each number of objectives 10 populations were drawn.

\begin{figure}[!hbp]
  \centering
  \includegraphics[scale=0.55,angle=0]{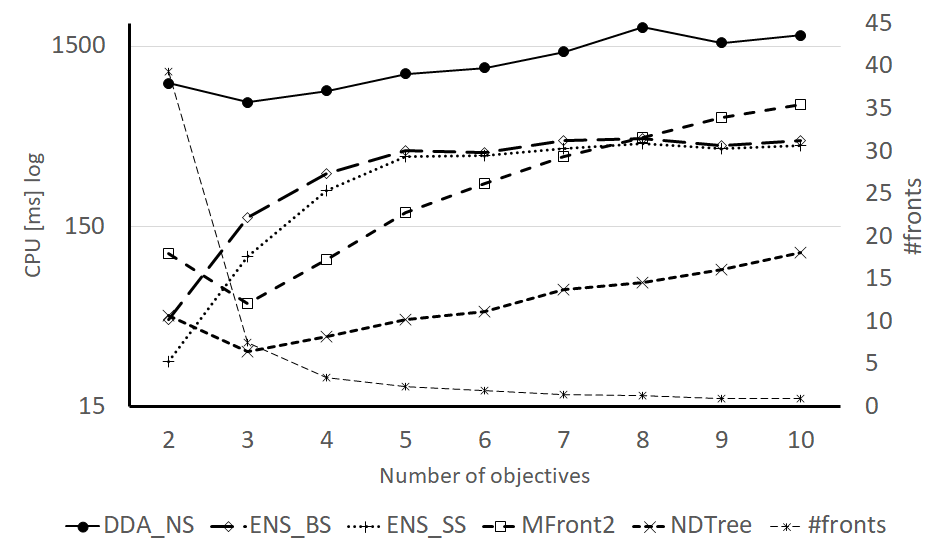}
   \caption{CPU times of non-dominated sorting algorithms for artificial sets with quality q3.}
   \label{fig:nds_our_cpu}
\end{figure}

\begin{figure}[!hbp]
  \centering
  \includegraphics[scale=0.55,angle=0]{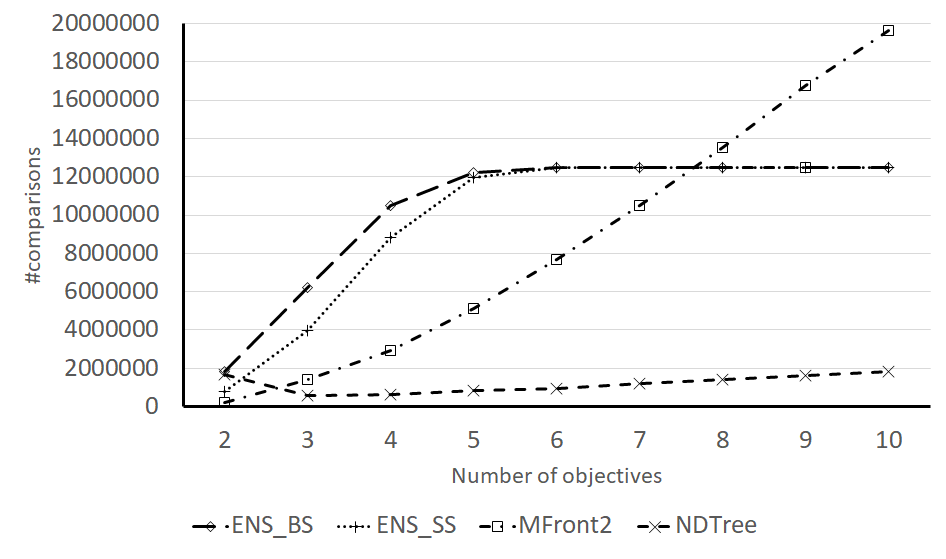}
   \caption{\#comparisons in non-dominated sorting algorithms for artificial sets with quality q3.}
   \label{fig:nds_our_comp}
\end{figure}

\begin{figure}[!hbp]
  \centering
  \includegraphics[scale=0.55,angle=0]{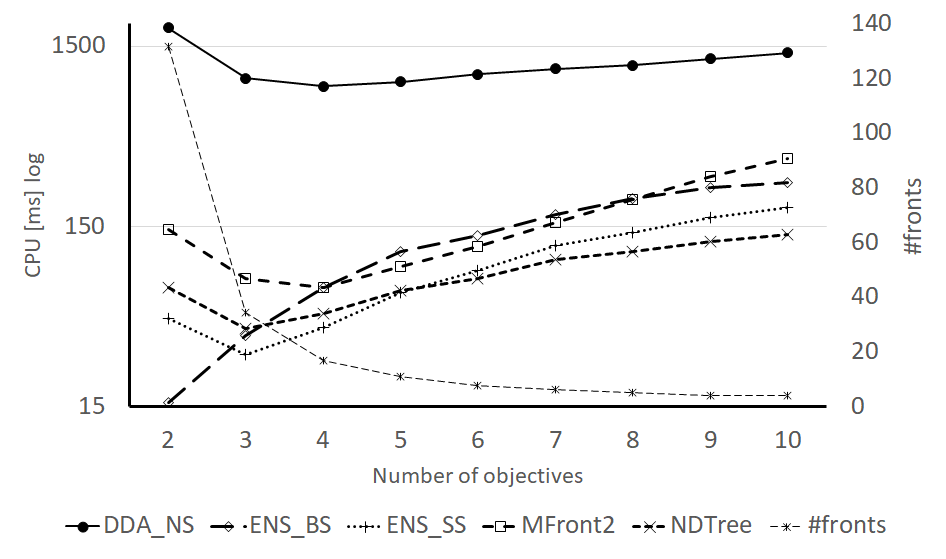}
   \caption{CPU times of non-dominated sorting algorithms for random populations.}
   \label{fig:nds_unif_cpu}
\end{figure}

\begin{figure}[!hbp]
  \centering
  \includegraphics[scale=0.55,angle=0]{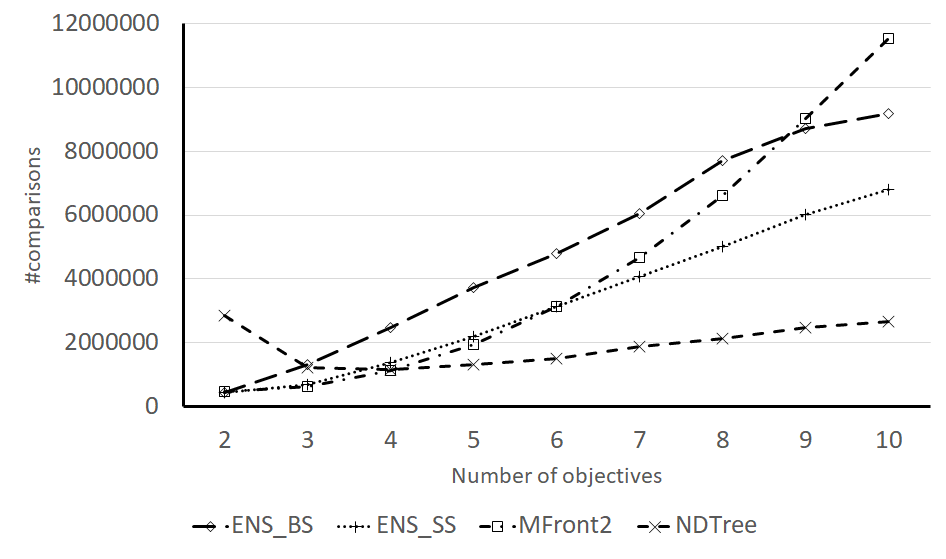}
   \caption{\#comparisons in non-dominated sorting algorithms for random populations.}
   \label{fig:nds_unif_comp}
\end{figure}

The results are presented in Figures \ref{fig:nds_our_cpu} to \ref{fig:nds_unif_comp}. We report both CPU times and the number of comparisons of points. We also show the number of fronts on the right $y$-axis. Please note that we do not show the number of comparisons for DDA-NS, since this algorithm does not explicitly compares points with the dominance relation. The slowest algorithm is DDA-NS. Our results are quite contradictory to the results reported in \cite{Zhou2017} where DDA-NS is the fastest algorithm in most cases. Please note, however, that in \cite{Zhou2017} the algorithms were implemented in MATLAB which, as the authors note, is very efficient in matrix operations. On the other hand, the CPU times reported in \cite{Zhou2017} are of orders of magnitude higher compared to our experiment which suggests rather that the MATLAB implementation of M-Front and ENS-BS/SS is quite inefficient. 

ENS-BS/SS performs well for populations with many fronts but its performance deteriorates when the number of fronts is reduced. As it was mentioned above if there is just one front in the population, the method boils down to the dynamic non-dominance problem solved with the simple list. This is why in the case of populations drawn from our sets which contain points of higher quality than random populations, and for higher numbers of objectives, where the populations often contain just one front, the number of comparisons saturates at a constant level.

In most cases ND-Tree-based non-dominated sorting is the most efficient method in terms of both CPU time and the number of comparisons. These results are very promising but only preliminary and further experiments, especially with populations generated by real MOEAs, are necessary. Please note that as suggested in \cite{Drozdik2015} the practical efficiency of the non-dominated sorting in the context of a MOEA may be further improved by maintaining the first front in a Pareto archive. 

\section{Conclusions}

We have proposed a new method for the dynamic non-dominance problem. According to the theoretical analysis the method remains sub-linear with respect to the number of points in the archive under mild assumptions and the time of processing a single point observed in the computational experiments are almost constant. The results of computational experiments with both artificial data sets of various global shapes, as well as results with sets of points generated by two different multiobjective methods, i.e. MOEA/D evolutionary algorithm and Pareto local search metaheuristic, indicate that the proposed method outperforms competitive methods in the case of three- and more objective problems. In biobjective case the best choice remains the sorted list.

We believe that with the proposed method for updating a Pareto archive, new state-of-the art results could be obtained for many multiobjective problems with more than two objectives. Indeed, results of our computational experiment indicate that the choice of an appropriate method for updating the Pareto archive may have crucial influence on the running time of multiobjective evolutionary algorithms and other metaheuristics especially in the case of higher number of objectives.

Interesting directions for further research are to adapt ND-Tree to be able to deal with archives of a relatively large but bounded size or to solve the static non-dominance problem. 

We have also obtained promising results by applying ND-Tree-based update to the non-dominated sorting and comparing it to some state-of-the-art algorithms for this problem. Further experiments are, however, necessary especially with populations generated by real MOEAs. Furthermore, good results of ENS algorithm for populations with many fronts suggest that a combination of ENS with ND-Tree may be an interesting direction for further research.

\section*{Acknowledgment}

The research of Andrzej Jaszkiewicz was funded by the the Polish National Science Center, grant no.~UMO-2013/11/B/ST6/01075.
  \vspace{-0.1cm}
\bibliographystyle{IEEEtran}
\bibliography{MaBiblio}

\end{document}